\documentclass[journal]{new-aiaa}
\usepackage{nccfoots}
\usepackage[utf8]{inputenc}
\usepackage{textcomp}
\usepackage{graphicx}
\usepackage{amsmath}
\usepackage[version=4]{mhchem}
\usepackage{longtable,tabularx}
\usepackage{makecell}

\title{On the Potential of Extending Aircraft Service Time Using Load Monitoring}

\author{Simon Pfingstl\footnote{Research Associate, Laboratory for Product Development and Lightweight Design, Boltzmannstr. 15, simon.pfingstl@tum.de, orcid=0000-0002-7282-889X, Corresponding author with shared co-first authorship.}}

\affil{Technical University of Munich, 85748 Garching, Germany}

\author{Dominik Steinweg\footnote{Research Associate, Economics and Transportation, Willy-Messerschmitt-Str. 1, dominik.steinweg@bauhaus-luftfahrt.net, orcid=0000-0002-0472-5573,  Corresponding author with shared co-first authorship. Both authors have contributed equally. The listed order of the two first authors was determined by rolling a dice.}}
\affil{Bauhaus Luftfahrt e. V., 82024 Taufkirchen, Germany}

\author{Markus Zimmermann\footnote{Professor, Laboratory for Product Development and Lightweight Design, Boltzmannstr. 15}}

\affil{Technical University of Munich, 85748 Garching, Germany}

\author{Mirko Hornung\footnote{Executive Director, Research and Technology, Willy-Messerschmitt-Str. 1}}
\affil{Bauhaus Luftfahrt e. V., 82024 Taufkirchen, Germany}

\begin{document}

\Footnotetext{}{Copyright \textcopyright$\,$2021 by Simon Pfingstl, Technical University of Munich, Dominik Steinweg, Bauhaus Luftfahrt e. V., Markus Zimmermann, Technical University of Munich, Mirko Hornung, Bauhaus Luftfahrt e. V. Published by the American Institute of Aeronautics and Astronautics, Inc. with permission.}

\maketitle

\begin{abstract}
\noindent Aircraft structures experience various kinds of loads over their entire lifetime, typically leading to fatigue and ultimately structural failure. In order to avoid structural failure during operation, the maximum number of flight cycles and flight hours is regulated by laws ensuring continued airworthiness. However, since every flight impacts the aircraft differently, not all airframes are equally stressed at the time of decommissioning. This paper proposes a new retirement criterion based on the so-called fatigue damage index (FDI). The criterion takes into account that aircraft are differently operated and thus enables an individual decommissioning of aircraft without compromising its safety. Based on aircraft sample data covering $95\%$ of the Airbus A320 fleet over two years, the enhanced decommissioning criterion is estimated to significantly extend the average aircraft service life. The impact varies within the fleet, depending on the experienced seat load factors, cruise altitudes, and taxi times considered for the individual aircraft during operation. While seat load factors and flight altitudes significantly affect the defined FDI, the influence of taxi times is only minor. Based on the estimated increase in aircraft service life, the paper argues that for service life extensions, the FDI shall be considered as the limit of validity in the regulatory framework governing the decommissioning of aircraft.
\end{abstract}



\section{Introduction and Motivation}
\noindent Ever increasing competition puts pressure on airlines to increase productivity in order to be financially successful \cite{Tretheway.2014}. In this regard, airlines can be challenged by a limited aircraft lifetime \cite{IATA.Workshop}. Regulatory guidelines set by the manufacturer and aviation authorities prohibit the operation of airframes exceeding a predefined number of flight cycles and flight hours, regardless of the actual structural condition. As a consequence, replacement aircraft are required to continue service while at the same time, the market value of decommissioned aircraft is reduced to scrap value. Therefore, the work at hand argues for a Fatigue Damage Index (FDI) to predict the necessity of aircraft decommissioning, advancing current regulations based on flight cycles and flight hours, without decreasing the level of safety. Utilizing available data, the resulting increase in aircraft lifetime through a better representation of operational usage and fatigue mechanisms is estimated at fleet level.
A brief introduction into the background of limited aircraft operating lifetimes and existing approaches to extend service goals is provided in the following sections. Subsequently, Chapter 2 describes the utilized data sets and approximations used in the work at hand. In Chapter 3, the computation of the FDI and its dependencies are explained. The resulting gain in aircraft lifetime, given in flight cycles and flight hours available at fleet level, is presented in Chapter 4. Following a discussion of the results in Chapter 5, the work is summarized in Chapter 6. 

\subsection{Motivation for Limited Aircraft Operating Lifetime}
\noindent On April 28, 1988, widespread fatigue and corrosion damage led to a structural failure and explosive decompression on Aloha Airlines flight 243 \cite{Airlines.}. In the aftermath, fatigue of aging metallic airframes, more specifically Widespread Fatigue Damage (WFD) was identified among the primary causes of this accident \cite{Schmidt.}. Since then, various “actions have been taken to ensure the continued structural airworthiness of the aging commercial transport aircraft fleet” \cite{Collins.1999} and to predict such phenomena \cite{harris1998analytical}. In this context, different revisions have been published, see, e.g., \cite{jones2014fatigue} where the author presents and discusses the damage tolerant approach by modeling crack growth, or \cite{tavares2017overview} for an overview of the regulatory evolution. To better address the decrease in ultimate strength of aging aircraft, a limit of validity for fatigue management programs of all future airplanes, based on full-scale fatigue testing, was established and incorporated into regulatory frameworks \cite{Eastin.2009}. This approach has been adopted by the Federal Aviation Administration (FAA) with 14 CFR § 129.115—“Limit of validity” and the European Union Aviation Safety Agency (EASA) with CS 25.571. The limit of validity for full-scale fatigue test data representing fatigue mechanisms must be defined in terms of flight cycles and flight hours for each aircraft type. On the Airbus A320 for example, the Design Service Goal (DSG) is set to 48,000 flight cycles and 60,000 flight hours. Over the course of the program, Airbus extended the original DSG of the Airbus A320 to the Extended Service Goal (ESG) by means of additional full scale fatigue tests to 60,000 flight cycles and 120,000 flight hours.

\subsection{Approaches Extending Aircraft Service Goals}
\noindent Considering that introducing predefined operating limits or service goals prohibit the use of aging aircraft that may still generate profits, multiple lifetime extension approaches are discussed in the literature and exist in practice. Also computation methods that support the decision process of aircraft retirement were presented, specifically in the military context \cite{lincoln1999economic}. The limit of validity is based on the initial full-scale fatigue test of the airframe. Additional tests covering the extended operating period may be performed to extend the limit of validity. However, full-scale fatigue tests are expensive and WFD is only limited to a known set of components, rather than the entire airframe \cite{Schmidt., Collins.1999}. Therefore, local structural health monitoring by means of dedicated sensors provide a tool for continuously monitoring an area of interest, mitigating the risks connected to fatigue and thus enable extended service goals \cite{Beral.}. This approach can already be found in practice. Amabile and Giacobbe (1991) converted strains measured by strain gauges applied to the trainer version of the military AMX jet aircraft to stress histories used to obtain load sequences \cite{AmabilePandGiacobbeT.1991}. These sequences are analyzed and synthesized using a rainflow cycle counting procedure \cite{Staszewski.2004}. The dominant parameters in any fatigue analysis are the magnitude and the mean value of the loads \cite{engle1985impact}. Therefore, researchers aimed to predict loads by different models, such as multiple regression, artificial neural networks, and reduced order models for helicopters \cite{haas1994identification,cook1994artificial} and aircraft \cite{castellani2015reduced,castellani2016parametric,castrichini2018gust}. The predicted loads could be used for probabilistic digital twin frameworks, as explained in \cite{li2017dynamic}, to update fatigue life predictions and to better schedule inspections (see \cite{ling2017information}). Such a digital twin concept can additionally consider other types of sensor signals, e.g., guided wave responses \cite{seshadri2017digitaltwin}.

Some load monitoring studies have been carried out at real aircraft structures \cite{lee2012review, HuntS.2001, Branco.2016}. For example, the Eurofighter Typhoon is equipped with a loads monitoring system that also considers strain gauge data \cite{HuntS.2001, boller2001ways}. Recordings of strain sequences at discrete locations can be incorporated into a loads model obtainable through extensive FE analysis, which allows stresses and strains at any location in the structure to be estimated. For the F-5E fighter aircraft of the Brazilian Air Force, fatigue monitoring at critical locations to increase maintenance efficiency is realized by measuring aircraft mass, flight speed, altitude, flap position, vertical load factor, and rolling speed \cite{Branco.2016}. Furthermore, the use of information from damage diagnosis and prognosis has been suggested to be incorporated into the aircraft control system to increase safety by limiting the flight envelope \cite{Seshadri.2017}. However, these load models are still rarely used directly for health monitoring in commercial aviation since they are expensive to obtain and update \cite{Staszewski.2004}. Further, monitoring system weight increases fuel burn and thus decreases the benefit of lifetime extensions, possibly to a point where costs surpass the benefits \cite{Beral.}. Alternatively, this drawback can be avoided by deriving the state of structural fatigue from available operational aircraft data. To this end, a dynamics-based loads monitoring algorithm has recently been suggested, which is capable of running in real-time on available computer hardware found on commercial aircraft, improving the knowledge about experienced loads in service \cite{Ntourmas.2020}. A similar approach for the synthesis of in-flight strains based on flight parameters using artificial neural networks has been suggested for military fighter aircraft \cite{Sharan.2013}. Thus, the limit of validity can be increased by improving the representation of fatigue while avoiding additional weight and costs. Co-opting sensors built into the aircraft for the purpose of operational load monitoring is advantageous because neither overall reliability nor complexity of the aircraft is negatively affected by additional sensors to be built in \cite{Staszewski.2004}. The health monitoring system for the Panavia Tornado was introduced as early as 1987 and 1988 by Bauer and Krauss, where they presented an exceedance monitoring with hard landing detection, speed exceedance, over-G and engine surge \cite{Bauer.1987, Krau.1988}. A similar sensor-based approach with hard landing detection and limit load exceedance was implemented for the Airbus A320 \cite{LaddaV.andMeyerH.J..1991}. In contrast, this paper argues for a method to compute the fatigue damage state of an aircraft by comparing actual loads to design loads using available information about its operation. 

\subsection{Proposing a Fatigue Damage Index as Aircraft Service Goal}
\noindent To increase the usable time of airframes, a shift from limits given in flight cycles and flight hours to a FDI is proposed. The representation of structural health can be expanded by additionally considering individual fatigue drivers rather than only relying on the number of flight cycles and flight hours alone, to better evaluate the operational environment and history of each individual aircraft. Combining this knowledge in an FDI enables an intuitive representation of the degree of structural fatigue and thus remaining useful lifetime.

In the work at hand, the plausibility of this approach is demonstrated and the potential of increased aircraft service lifetimes estimated. Therefore, an approach to calculate the FDI for both the wing and fuselage is introduced based on previous work described in the literature. The impact of current and proposed decommissioning criteria on aircraft lifetimes is evaluated using sample data of Airbus A320 fleet movements. Parameters required for the computation of the FDI not contained in the available sample data, including flight altitude and aircraft weight, are estimated using simplified models.

\section{Data Set and Approximations}
\noindent Considered Airbus A320 fleet operations are based on ADS-B data provided by Flightradar24. A total number of 7,951\footnote{On July 25, 2019, a total of 8,316 Airbus A320 were operational \cite{airlinerlist}.} aircraft were covered over a period of 797 days between September 23, 2017 and November 29, 2019, amounting to a total of 16,625,103 flights. The data set contains an average number of 2.62 flights per aircraft and day, thus covering only a fraction of all flights. The usage behavior of the fleet is thus extrapolated, assuming that the available data is representative for every individual aircraft. The payload on each flight is approximated by the average seat load factor on a specific origin-destination pair for a single airline, based on leg data provided by Sabre. The average taxi time is considered by continent, based on data made available by Eurocontrol. Available usage data is replicated until the aircraft is retired, assuming unchanged usage behavior over the entire aircraft lifetime. Selected operational characteristics of the entire fleet are provided in Fig. \ref{FIG:1}. Properties of the aircraft considered in this work are summarized in Table \ref{Airbus Performance Characteristics}. Differences between variants and versions are neglected. Within the scope of this work, the actual takeoff weight and top of descent on each flight is not available and thus approximated as follows:

\begin{figure*}
	\centering
	\includegraphics[scale=.75]{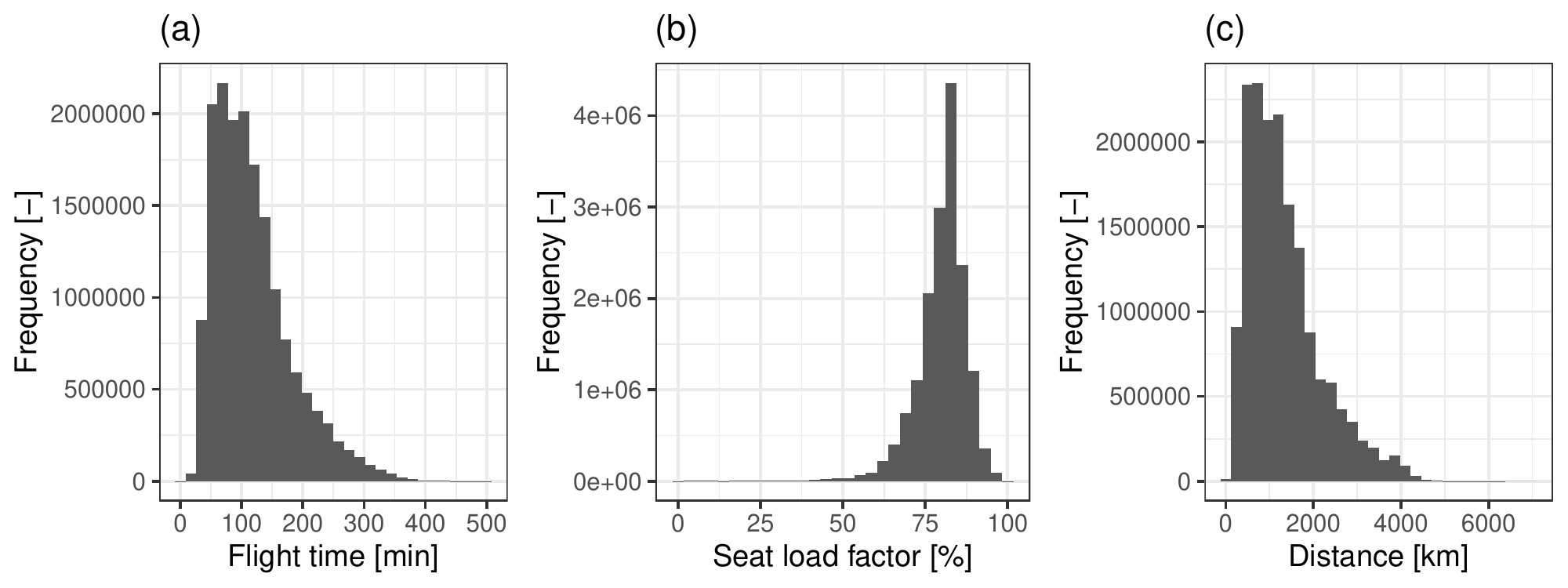}
	\caption{Average flight time (a), seat load factor (b) and flight distance (c) in considered Airbus A320 fleet.}
	\label{FIG:1}
\end{figure*}

\begin{table}
	\caption{Assumed Airbus A320-100 performance characteristics \cite{AIRBUSS.A.S..2019, Roux.2007}. Performance variations across Airbus A320 family are neglected.}\label{Airbus Performance Characteristics}
	\centering
	\begin{tabular*}{9.5 cm}{llll}
		\hline
		\textbf{Performance characteristics} & \textbf{Size} \\
		\hline
		Additional fuel reserve [$W_{Fuel,res}$] 	& 2,500 kg \\					
		Cruise speed at FL 390 [$M_c$] 				& 0.79 \\		
		Fuel consumption during taxi [$F_{Taxi}$] 	&	0,1 kg/s \\				
		Fuel reserve factor [$F_{Fuel,res}$]		& 1.05 \\				
		Gravity constant [$g$] 						& 9.81 m/$\mathrm{s^2}$ \\
		Lift-to-drag ratio [$\frac{L}{D}$] 			& 15 \\			
		Maximum payload [$W_{PL,max}$]     		 	& 16,600 kg \\			
		Maximum takeoff weight (MTOW) [$W_{MTOW}$] 	& 73,500 kg \\				
		Maximum weight of fuel [$W_{Fuel,max}$] 	& 20,000 kg \\					
		Minimum payload (Crew, e.g.) [$W_{misc}$] 	&	600 kg \\				
		Minimum weight of fuel [$W_{Fuel,prop,min}$]& 5,000 kg \\						
		Number of seats [$N_{Seats}$] 				&	160 \\	
		Operating empty weight [$W_{OEW}$]     		& 42,200 kg \\				
		Seat Load factor ($N_{Seats,occupied}$) [$LF$] & $\mathrm{0 - 100} \%$ \\	
		Specific fuel consumption [$SFC_{cruise}$]	& 16.88 $\mathrm{g/kN \cdot s}$ \\
		Speed of sound at FL 390 [$a$] 				& 295 m/s \\		
		Weight per passenger [$W_{PAX}$] 			& 100 kg \\			
		\hline
	\end{tabular*}
\end{table}

\subsection{Takeoff Weight}
\begin{figure}
	\centering
	\includegraphics[scale=.62]{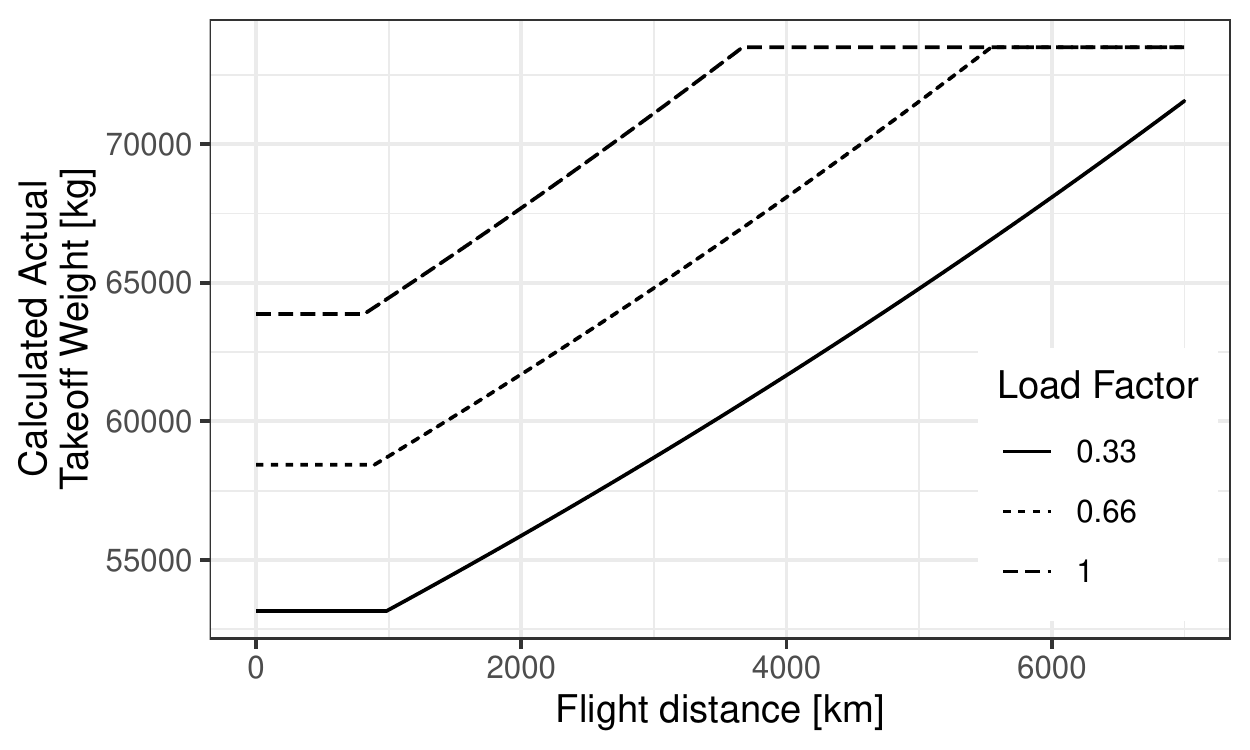}
	\caption{Calculated aircraft takeoff weight for a given flight distance assuming selected seat load factors and considering fuel consumption.}
	\label{range_takeoff_weight}
\end{figure}

\noindent Fuel weight and consequently takeoff weight on each flight is calculated as follows:
\begin{equation}
  W_{Taxi Fuel,dest}  = T_{Taxi,dest} \cdot F_{Taxi},
\end{equation}
where $T_{Taxi,dest}$ is the required taxi time at the destination and $W_{Taxi Fuel,dest}$ the weight of fuel required at the destination for taxiing. The resulting aircraft takeoff weight without fuel for propulsion can be calculated by
\begin{equation}
W_{TOW w/o Fuel} =   W_{OEW} + W_{misc} + LF \cdot N_{Seats} \cdot W_{Pax} + W_{Taxi Fuel,dest}
\end{equation}
The weight of required fuel for the cruise flight is derived by using the Breguet range equation. To approximate the total fuel weight, a reserve $W_{Fuel,res}$ is considered to account for the planning of an alternate airport. Further, the total weight of fuel is increased by a factor $F_{Fuel,res}$ to consider the required fuel for take-off and holding patterns.
\begin{equation}
W_{Fuel,prop} = \left( W_{TOW w/o Fuel} +  W_{Fuel,res}  \right) \cdot \exp \left( \frac{ \frac{d_{flight}}{a \cdot M_c} \cdot g \cdot SFC_{cruise}  }{\frac{L}{D}} -1 \right) \cdot F_{Fuel,res}
\end{equation}
Additionally, airline procedures require fuel at least equal or exceeding a predefined minimum $W_{Fuel,prop,min}$ on take-off, considered by
\begin{equation}
W_{Fuel,prop}= max \left( W_{Fuel,prop}, W_{Fuel,prop,min}  \right).
\end{equation}
In the work at hand, $W_{Fuel,prop,min}$ is considered equal for all airlines. Further, the weight of the aircraft on takeoff is limited by the MTOW set by the manufacturer 
\begin{equation}
W_{TOW,Total}  = min \left(  W_{TOW w/o Fuel} + W_{Fuel,prop}, W_{MTOW} \right).
\end{equation}
The aim of this approach is to provide a fair proxy of the fuel weight in practice rather than calculate the exact fuel consumption during flight. The consequent takeoff weight over flight distance for varying seat load factors is illustrated in Fig. \ref{range_takeoff_weight}. 

\subsection{Maximum Altitude}
\begin{figure}
	\centering
	\includegraphics[scale=.62]{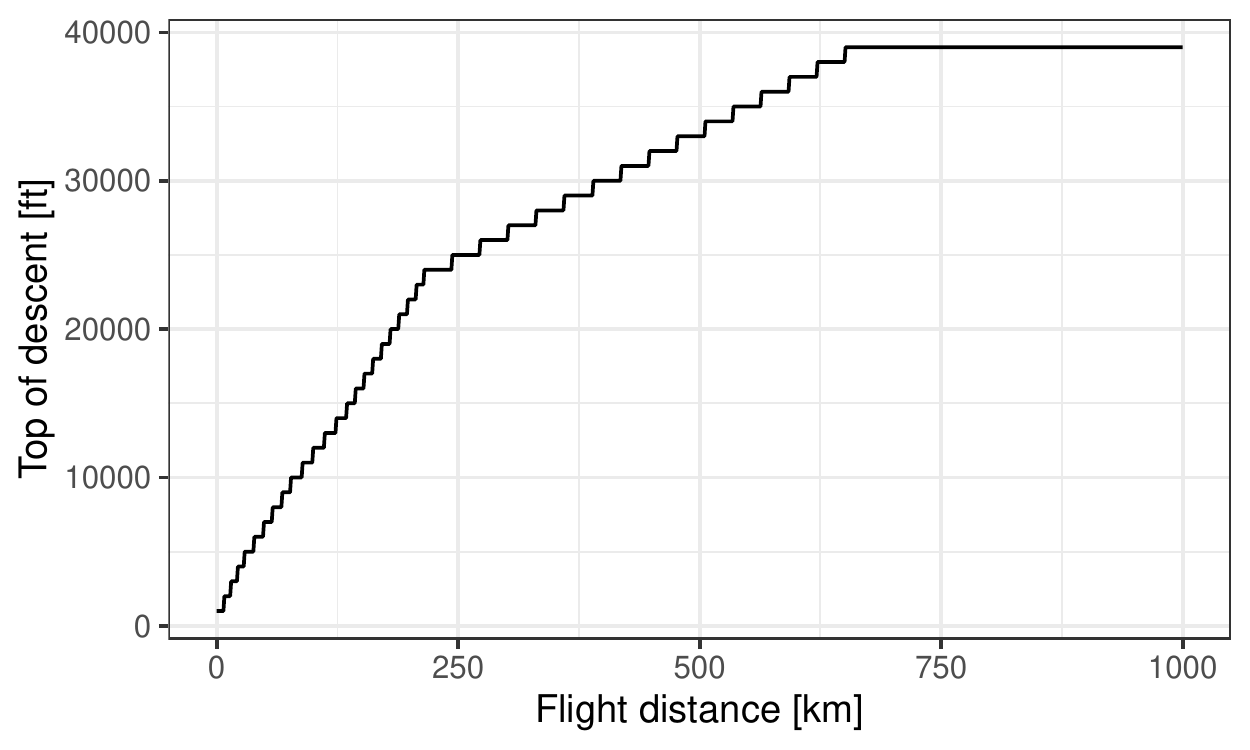}
	\caption{Maximum flight altitude assumed for a given flight distance. The aircraft climb performance is listed in Tab.~\ref{Considered Aircraft Performance} and based on \cite{Eurocontroll.2020}.}
	\label{range_altitude}
\end{figure}

\noindent The maximum altitude of every flight is based on the aircraft performance considered by Eurocontrol, summarized in Table \ref{Considered Aircraft Performance}. Depending on the distance between origin and destination, the aircraft climbs to the highest attainable flight level, neglecting the semicircular separation rule. The resulting maximum altitude over distance is given in Fig. \ref{range_altitude}.

\begin{table}
	\caption{Considered Aircraft Performance \cite{Eurocontroll.2020}.}\label{Considered Aircraft Performance}
	\centering
	\begin{tabular*}{8.5 cm}{llll}
		\hline
		\textbf{Flight Phase}		&	\textbf{Rate of Climb} & \textbf{Speed}	\\
		\hline
		Initial climb				&	2500 ft/min	&	IAS 175 kts	\\
		Climb to FL 150				&	2000 ft/min	&	IAS 290 kts	\\
		Climb to FL 240				&	1400 ft/min	&	IAS 290 kts	\\
		MACH climb to FL 390		&	1000 ft/min	&	MACH 0.78	\\
		Cruise						&	0 	 ft/min	&	MACH 0.79	\\
		Initial descent to FL 240  	& 	1000 ft/min	&	MACH 0.78	\\
		Descent to FL 100 	 		&	3500 ft/min	&	IAS 290 kt	\\
		Approach 	 				&	1500 ft/min	&	IAS 250 kt	\\
		\hline
	\end{tabular*}
\end{table}

\section{Fatigue Damage Index}
\noindent In order to account for different fatigue drivers and to approximate the remaining useful life of the airframe, standard fatigue life equations are applied. Since the load spectrum during a flight does not occur with a fixed stress ratio $R$, an approximation of the Haigh diagram \cite{Rennert.2012} is used in order to map the load spectrum to an equivalent load spectrum with $R =-1$ (see Eq. \ref{eq:Haigh1}-\ref{eq:Haigh4}). The stress amplitude $S_{a,R=-1}$ for $R =-1$ is therefore dependent on the mean stress $S_m$, the amplitude stress $S_a$ for a known $R \neq -1$, and the material constant $M_\sigma$. Subsections 3.1. and 3.2. describe the computation of $S_m$ and $S_a$.

\begin{equation}
\label{eq:Haigh1}
\begin{array}{ll}
	\text{for } R > 1 \text{:} \\
	S_{a,R=-1} = S_a (1-M_\sigma)
\end{array}
\end{equation}

\begin{equation}
\label{eq:Haigh2}
\begin{array}{ll}
	\text{for } -\infty \leq R \leq 0 \text{:} \\
	S_{a,R=-1} = S_a (1+M_\sigma \frac{S_m}{S_a})
\end{array}
\end{equation}

\begin{equation}
\label{eq:Haigh3}
\begin{array}{ll}
	\text{for } 0 < R < 0.5 \text{:} \\
	S_{a,R=-1} = S_a \frac{(1+M_\sigma)(3+M_\sigma \frac{S_m}{S_a})}{3+M_\sigma}
\end{array}
\end{equation}

\begin{equation}
\label{eq:Haigh4}
\begin{array}{ll}
	\text{for } R > 0.5 \text{:} \\
	S_{a,R=-1} = S_a \frac{3(1+M_\sigma)^2}{3+M_\sigma}
\end{array}
\end{equation}

\noindent After computing the stress amplitude for $R=-1$, the FDI, which describes the remaining useful life by a factor between 0 and 1 (where $\text{FDI}=1$ indicates no remaining useful life) can be evaluated by the corresponding S-N curve and Miner’s rule (Eq. \ref{eq:Miner}). Miner’s rule \cite{Miner.1945} assumes that the fatigue damage of different stress amplitudes accumulates linearly, where the fatigue damage is the ratio of the number of cycles of the applied load $n_i$ to the number of cycles $N_i$, which can be withstood by the structure according to the S-N curve.

\begin{equation}
\label{eq:Miner}
	\text{FDI}= \sum_i \frac{n_i}{N_i}
\end{equation}

\noindent To consider the different fatigue drivers, the aircraft is divided into two main structural parts: the wing and the fuselage. Within the work at hand, taxi time, payload, maximum flight altitude, and flight time are considered causes for fatigue damages. For both the wing and fuselage the aluminum Al 2024-T3 with $M_\sigma=0.4$ \cite{Haibach.2006} is assumed. Moreover, the corresponding S-N curve (Eq. \ref{eq:SNcurve}) with a stress concentration factor of $K_t=2.5$, which is usually found at holes, is used (S-N parameters: $C_1=63 \, \text{MPa}$; $C_2=470 \, \text{MPa}$; $C_3=3.50$; $C_4=2.07$ \cite{Handbuch.1978}). $C_1$ defines the fatigue limit, $C_2$ the ultimate strength, and $C_3$ and $C_4$ the slope of the S-N curve.

\begin{equation}
\label{eq:SNcurve}
	N_i = 10^{C_3 \left[ ln \left(\frac{C_2-C_1}{S_{a,i}-C_1} \right) \right]^{1/C_4}}
\end{equation}

\subsection{Fatigue Damage Index of the Wing}
\begin{figure}[h]
	\centering
		\includegraphics[scale=.12]{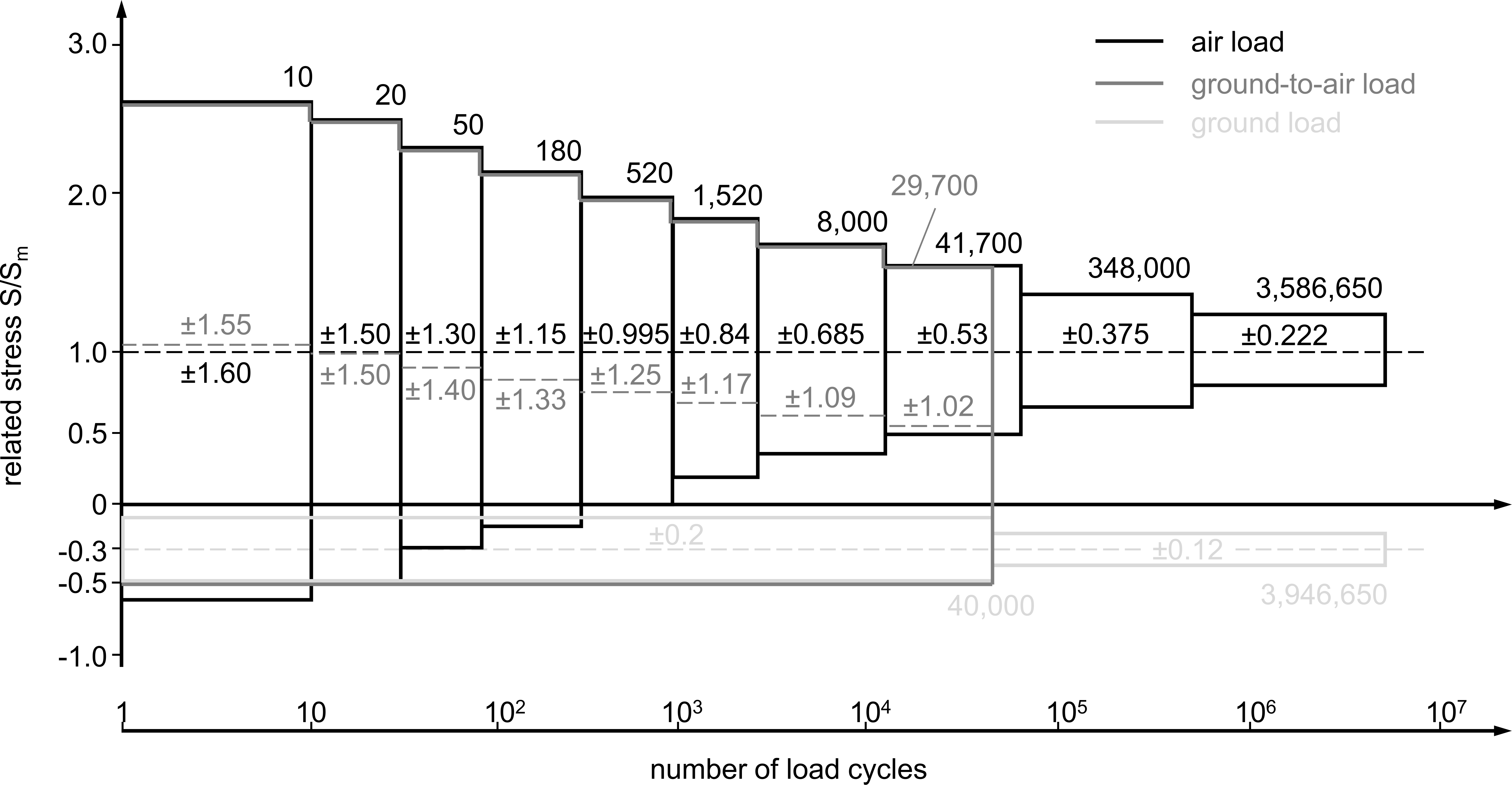}
	\caption{Transport Wing Standard for 40,000 flights based on \cite{schutz1973standardisierter}.}
	\label{fig:TWIST}
\end{figure}

\noindent In order to compute the FDI of the wing, the stress amplitudes and the mean stresses during flight have to be evaluated. The Transport Wing Standard (TWIST) shown in Fig. \ref{fig:TWIST} depicts the frequency distribution of the bending moment at a wing root of civil and military transport aircraft for 40,000 flights \cite{schutz1973standardisierter}. The TWIST was generated using measurements and computations and is divided into three parts: loads during flight, loads on the ground (due to taxi), and ground-to-air loads.

First, the fatigue life based on the design loads can be computed with Eq. \ref{eq:Haigh1}-\ref{eq:SNcurve}. The TWIST stresses are related to the mean stress, which is assumed to be $100 \, \text{MPa}$ for the design loads at a maximum takeoff weight (MTOW) $W_{MTOW}=73.5 \, \text{t}$, maximum payload $W_{PL,max}=16.6 \, \text{t}$, operating empty weight $W_{OEW}=42.2 \, \text{t}$, and maximum fuel weight $W_{fuel,max}=20 \, \text{t}$ for an Airbus A320 \cite{AIRBUSS.A.S..2019}. The corresponding numbers of cycles are shown in Fig. \ref{fig:TWIST}. Furthermore, the design flight time is set to $FH_{des}=2 \, \text{h}$ \cite{AIRBUSS.A.S..2019}, and the design taxi time to $T_{taxi,total,des}=25 \, \text{min}$ \cite{Eurocontroll.2020}. These times are assumed to lead to the numbers of cycles presented by the TWIST. Considering the taxi time rather than distance allows for a straightforward computation of fuel burn and weight changes during taxiing and further takes airport congestion into account. Due to the exploratory nature of the study, taxi distances and taxiway conditions, both leading to an airframe excitement and thus fatigue, are neglected.

Second, since both the takeoff weight $W_{TOW}$ (herein estimated through the aircraft seat load factor and fuel weight) and the flight time are known, relative changes can be taken into account for the fatigue computation. The paper at hand assumes following linear dependencies: The mean stress is linearly dependent on the aircraft weight, the number of cycles during flight is linearly dependent on the flight hours, and the number of cycles on the ground are reduced in proportion to the reduction of the taxi time. Now, the difference in the fatigue life represented by the change of the FDI due to monitoring can be evaluated.

\subsection{Fatigue Damage Index of the Fuselage}
\noindent The stresses in the fuselage result mostly from the difference between inner and outer pressure. Due to lift forces of the wing, the fuselage is also subjected to bending loads. However, in the present study, this influence is neglected. In practice, there can be complex interdependencies between assemblies that influence fatigue life. Therefore, cracks sometimes occur at supposedly unexpected locations, as seen for example on early Airbus A380 aircraft \cite{EASA-12-0013,EASA-19-116}. The proposed FDI criterion, however, shall give a rather global assessment of fatigue life. Since unexpected events can still occur, inspections or repairs cannot be omitted. Using this assumption, the tangential stress $S_{tan}$ and axial stress $S_{axial}$ present in the fuselage can be modeled by Barlow’s formula (see Eq. \ref{eq:Kesselformel}) with the differential pressure $\Delta p$, average of the fuselage diameter $d_m$, and sheet thickness $t$. Since the tangential stress is twice as great as the axial stress, only the tangential stress is considered in the following. The thickness is set to $t=1.0 \, \text{mm}$ and according to \cite{AIRBUSS.A.S..2019} the fuselage diameter is $d_m=4.14 \, \text{m}$. 
\begin{equation}
\label{eq:Kesselformel}
	S_{tan}=\frac{p \, d_m}{2 \, t}, \\
	S_{axial}= \frac{p \, d_m}{4 \, t}
\end{equation}
The differential pressure is mostly affected by the highest altitude reached on an individual flight. In the work at hand, this altitude is assumed to be only dependent on the distance between origin and destination as illustrated in Fig. \ref{range_altitude}. The dependency is modeled in Eq. \ref{eq:HeightPressure}, where $p_0=1013.25 \, \text{hPa}$ is the pressure at sea level, $H_0=8,435 \, \text{m}$, and $H$ is the flight altitude.
\begin{equation}
\label{eq:HeightPressure}
	p=p_0 \left( 1-e^{-\frac{H}{H_0}} \right)
\end{equation}
According to \cite{AIRBUSS.A.S..2019}, the maximum design flight altitude is set to $H_{max,des}=39,100 \, \text{ft}$. Given that the pressure results in a pure swelling load with $R=0$, the mean stress $S_m$ and the amplitude stress $S_a$ can be computed as (Eq. \ref{eq:Sm}):
\begin{equation}
\label{eq:Sm}
	S_m=S_a=\frac{S_{tan}}{2}
\end{equation}
Now the fuselage FDI corresponding to the design parameters can be computed with Eq. \ref{eq:Haigh1}-\ref{eq:SNcurve}. Since the actual flight altitude is known, it can be taken into account and the change of the FDI due to monitoring can be computed.

\subsection{Dependencies of the Fatigue Damage Index}
\begin{figure*}
	\centering
	\begin{tabular*}{20 cm}{ll}
		(a) & (b) \\
		\includegraphics[scale=.5]{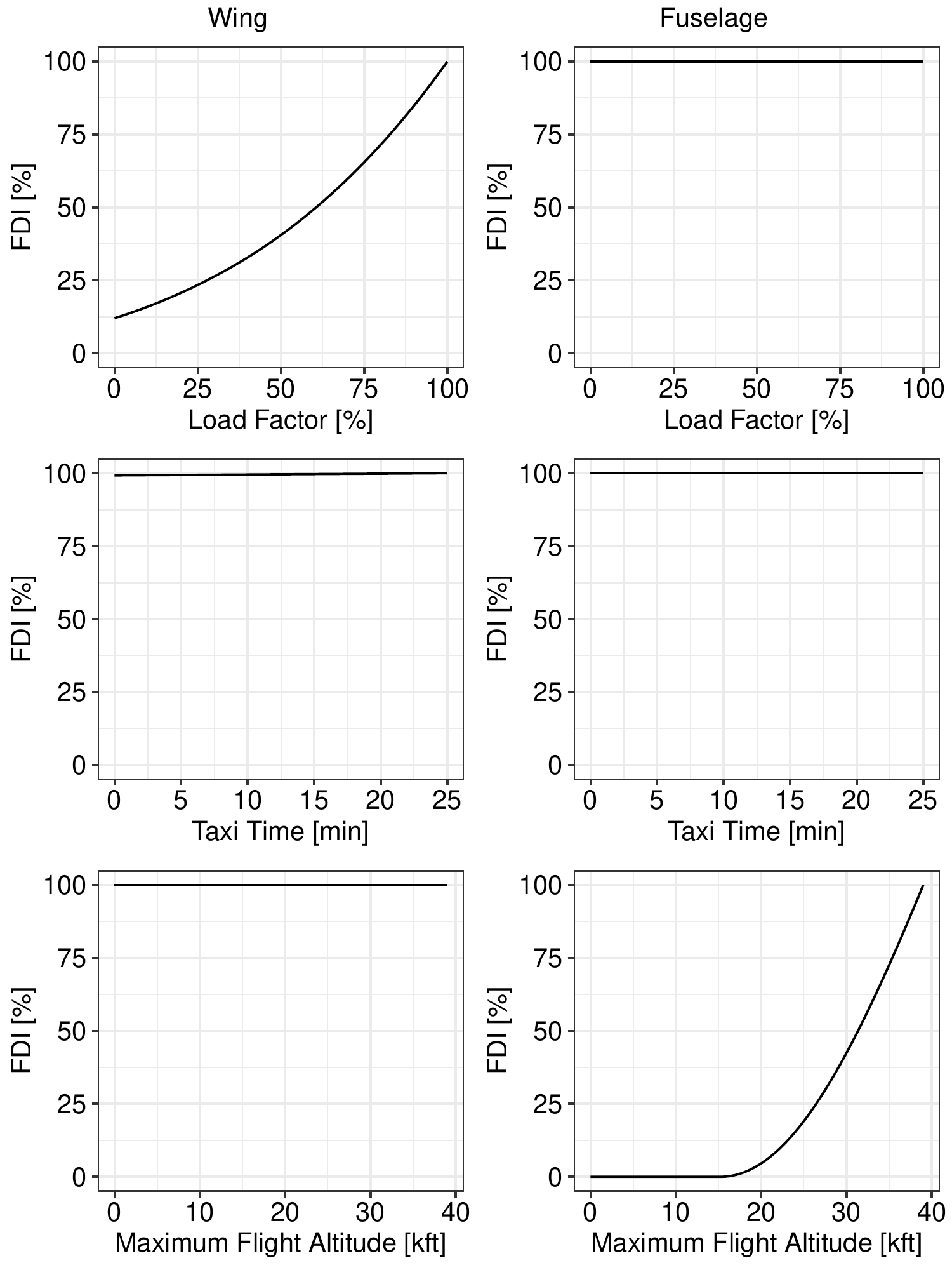} & \includegraphics[scale=.5]{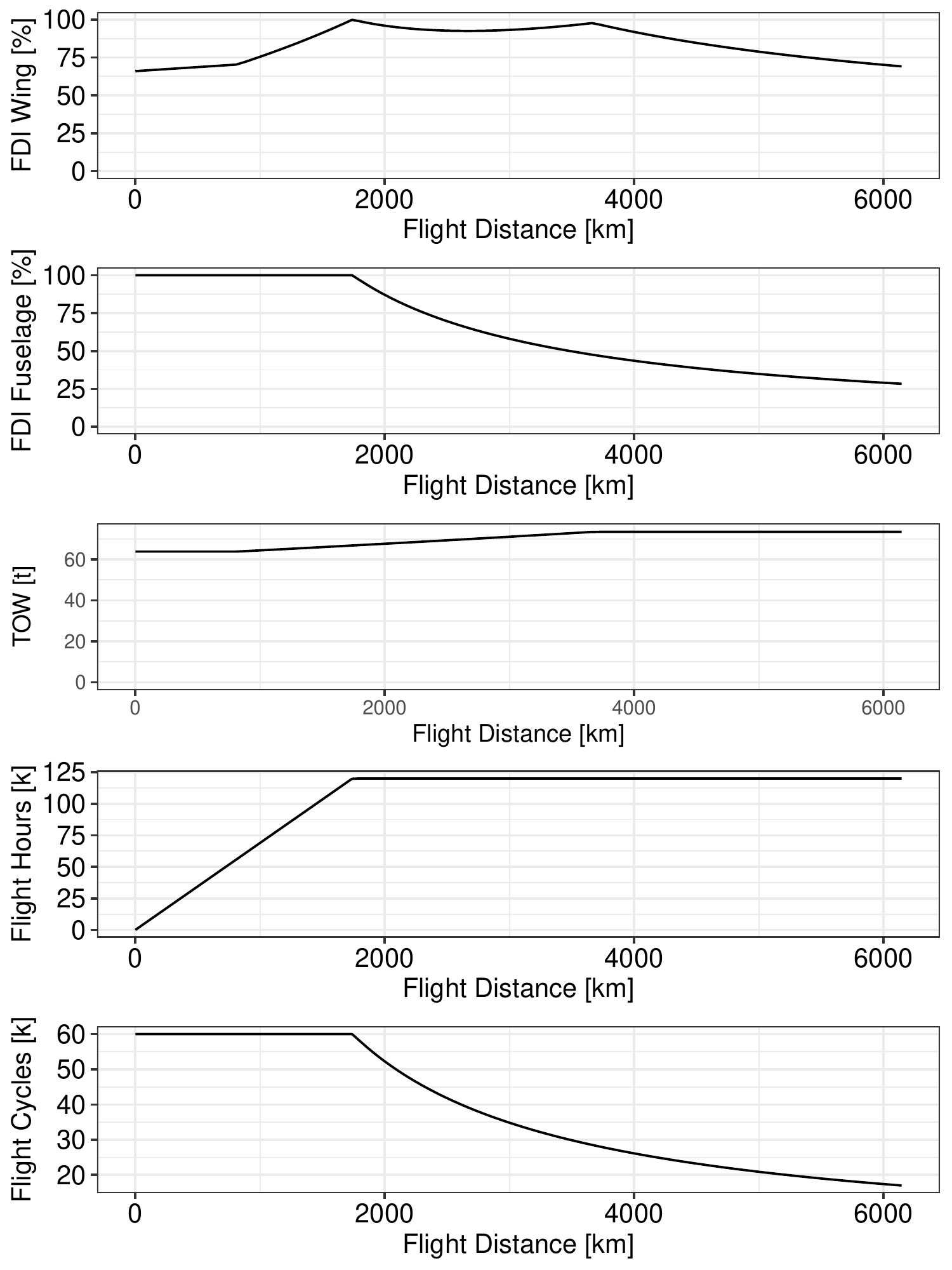}
	\end{tabular*}
	\caption{Evolution of the FDI (a), and influence of flight distance (b) for an A320 at ESG.}
	\label{fig:dependencies}
\end{figure*}

\noindent To clarify the dependencies of the FDI, Fig. \ref{fig:dependencies} illustrates the influences of its parameters. In all diagrams of Fig. \ref{fig:dependencies}a, only a single parameter is changed, ceteris paribus. The wing FDI shows a strong dependency on the seat load factor. Due to a larger mass, the mean stress intensifies, increasing the FDI. In contrast, the fuselage is not affected by the seat load factor since its FDI is completely driven by the flight altitude based on previously mentioned assumptions. The fuselage FDI is also independent of the taxi time, whereas the wing FDI shows a minor dependency on it because the taxi load's mean stress is negative and the amplitude stress is rather small (see Fig. \ref{fig:TWIST}). Furthermore, the wing FDI is assumed to be independent of the flight altitude, whereas the fuselage FDI shows a strong dependency on it.

Moreover, the influence of the flight distance is visualized in Fig. \ref{fig:dependencies}b, assuming that a single aircraft has been operated on a single flight distance for its entire life until the Extended Service Goal (ESG). As a result of limiting flight cycles and flight hours at predefined values, it can be seen that a longer flight distance than the considered average in the ESG leads to fewer flight cycles until decommissioning and therefore to smaller FDIs. Likewise, a shorter flight distance results in decreased weight of fuel and flight hours. Accumulating these shorter flights over the aircraft lifetime until the service goal is reached reduces the fatigue indices of both the wing through less weight and the fuselage through lower flight altitudes.


\begin{figure}
	\centering
	\begin{tabularx}{16cm}{p{8cm}l}
		(a) FDI & (b) FDI + taxi times \\
		\includegraphics[scale=.45]{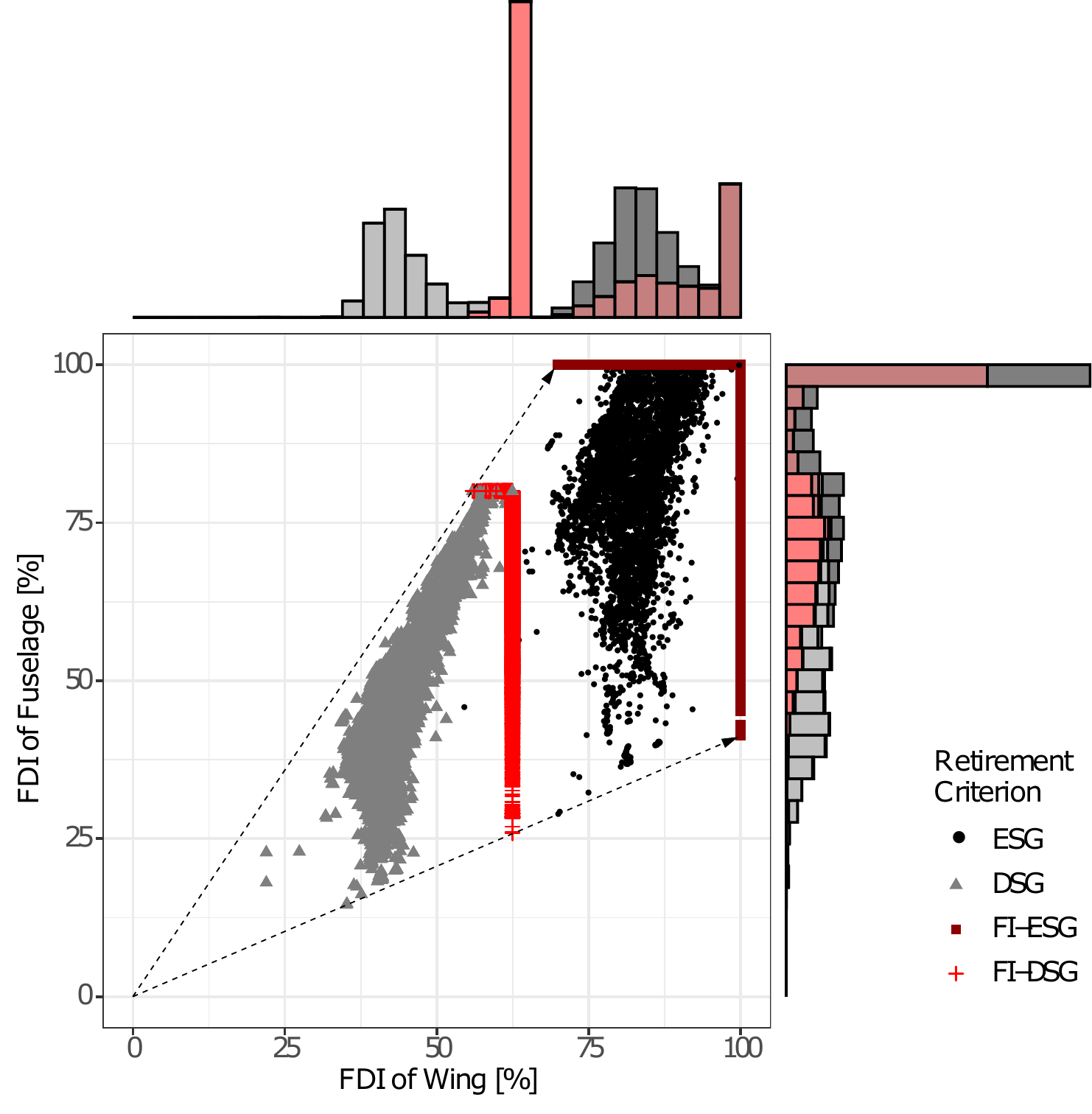} & \includegraphics[scale=.45]{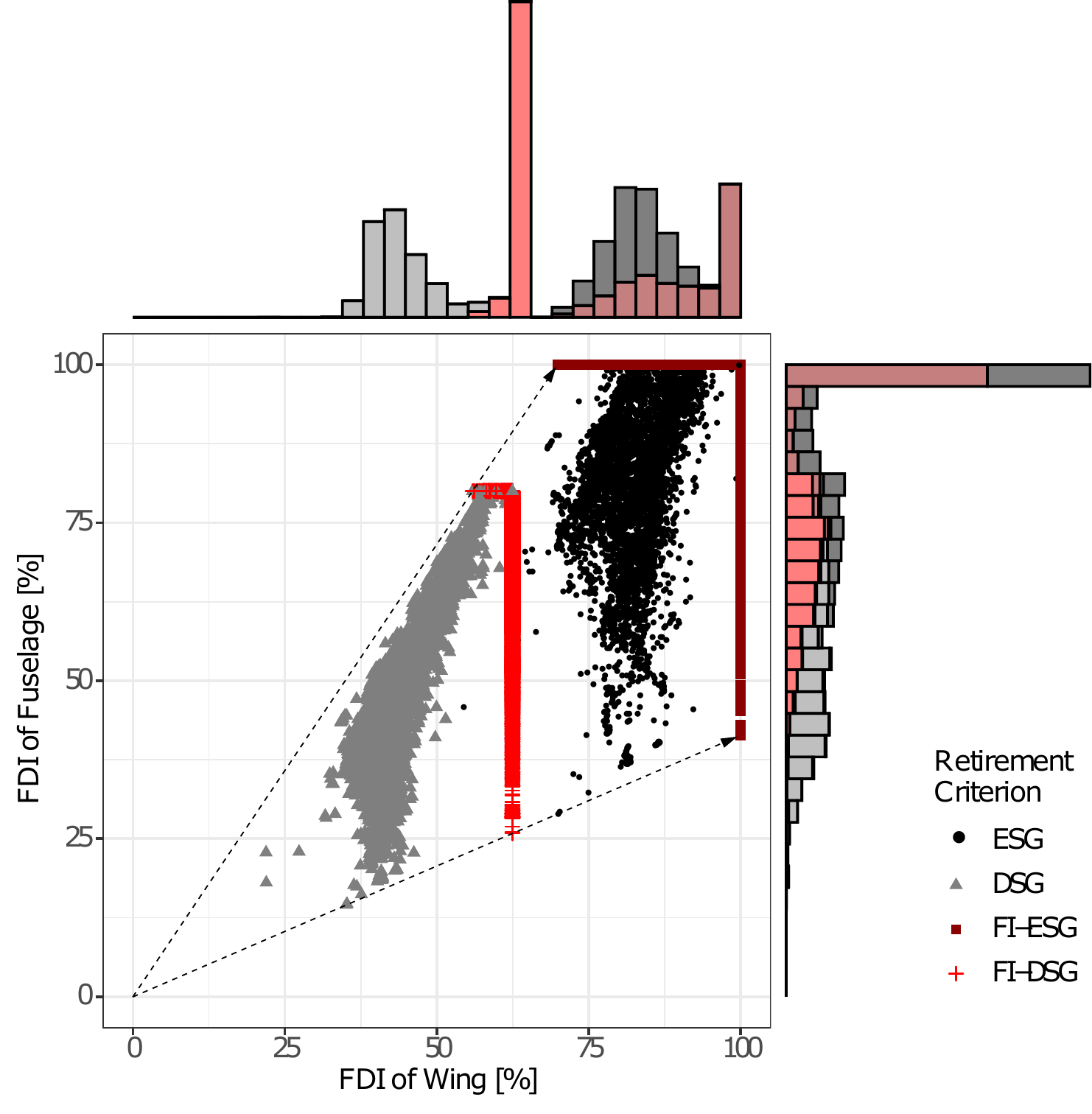} \\
		(c) FDI + maximum flight altitude & (d) FDI + seat load factor \\
		\includegraphics[scale=.45]{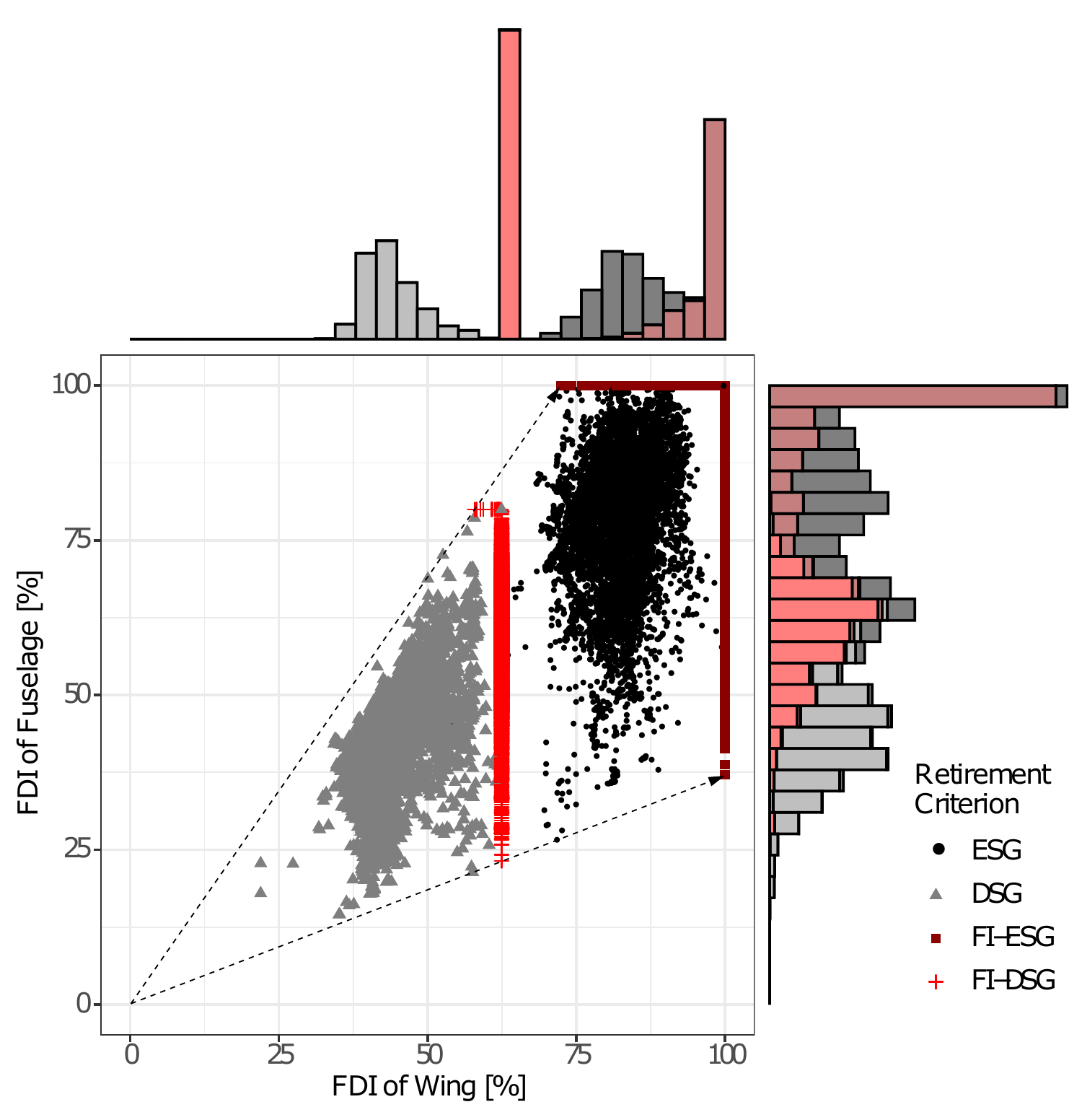} & \includegraphics[scale=.45]{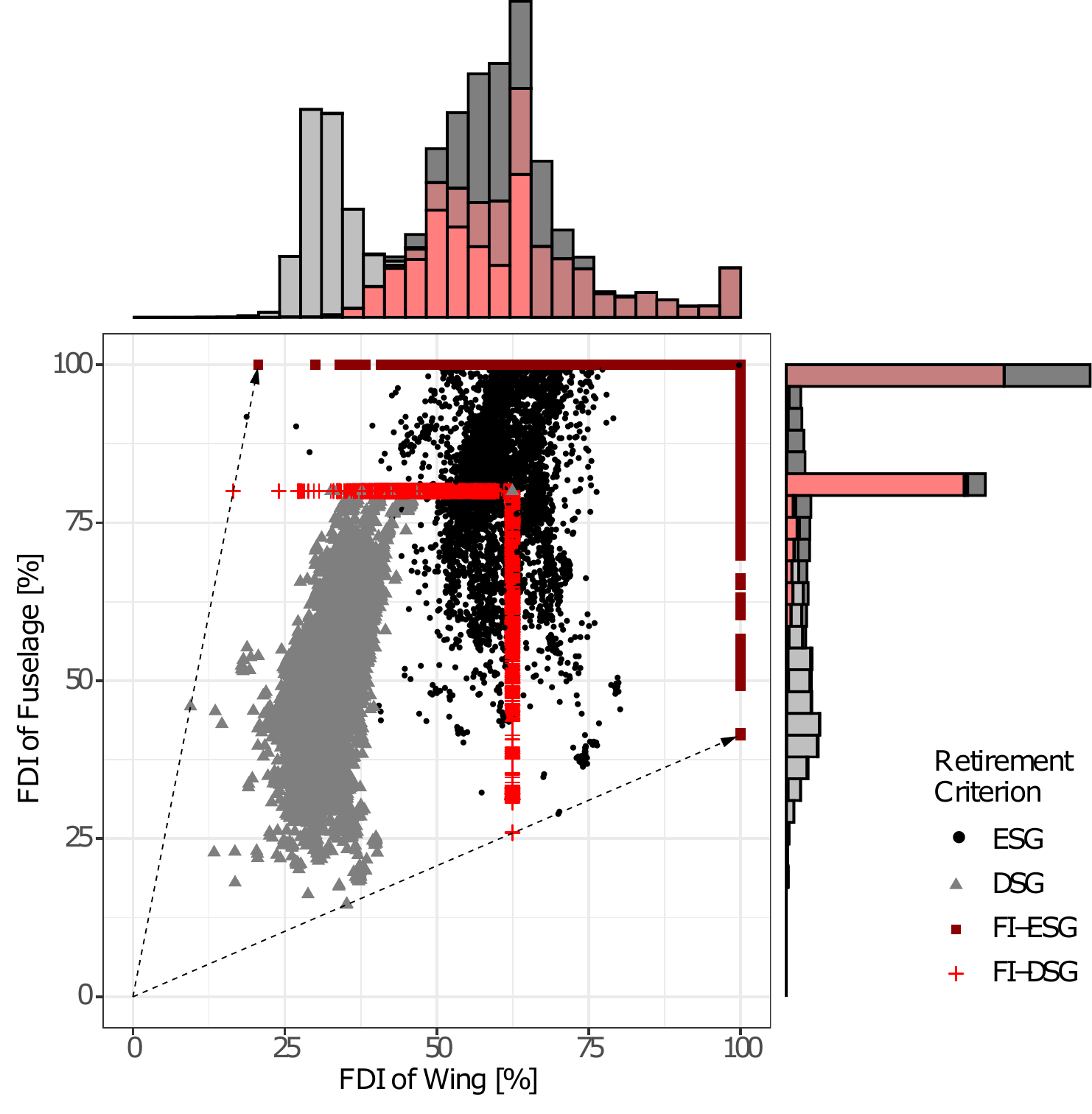} \\
		(e) FDI +  altitude and seat load factor & (f) FDI + altitude, seat load factor, and taxi time \\
		\includegraphics[scale=.45]{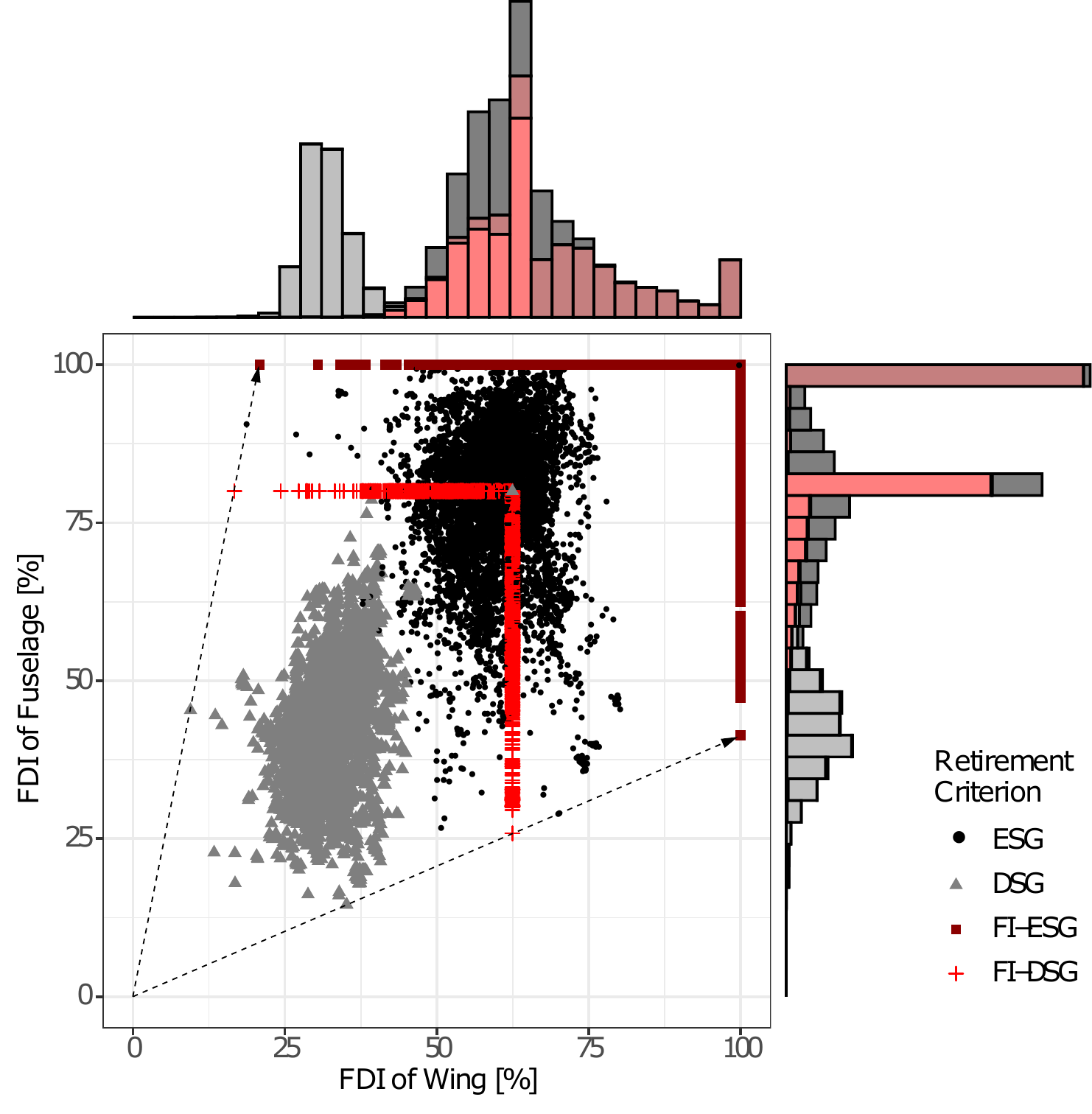} & \includegraphics[scale=.45]{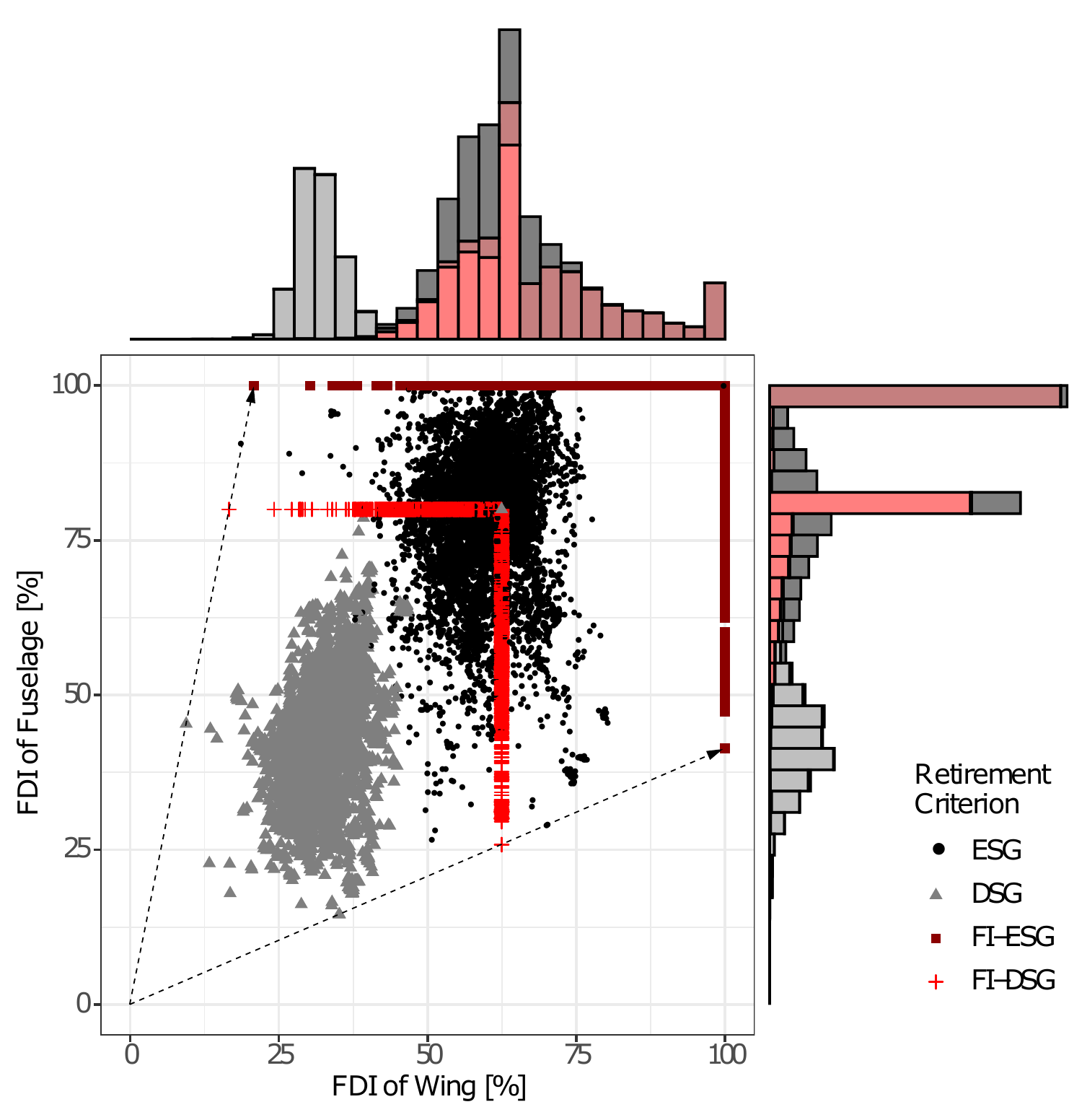}
	\end{tabularx}
	\caption{Computed FDIs at fleet level. The baseline is shown in (a). Additional fatigue drivers are considered as described in each individual sub-caption.}
	\label{fig:FDIs}
\end{figure}

\clearpage

\section{Results}

\noindent For this study, four scenarios are considered for the Airbus A320 fleet: First, the FDI is evaluated until the aircraft has to be retired due to the DSG at 48,000 flight cycles and 60,000 flight hours (gray points, Fig. \ref{fig:FDIs}). Second, the fleet is retired at the ESG of 60,000 flight cycles and 120,000 flight hours (black points, Fig. \ref{fig:FDIs}). Third, the fleet is operated until every aircraft reaches the FDI equaling that of an aircraft operated until DSG under design loads (bright red points, Fig. \ref{fig:FDIs}). And fourth, every aircraft is operated until its FDI reaches a level corresponding to design loads at ESG (dark red points, Fig. \ref{fig:FDIs}). The FDI is computed for both the wing and the fuselage for a total of 7,951 aircraft and is set equal to 100$\%$ for an aircraft operated at design loads until the ESG (assuming an average flight time of 2 h with a seat load factor of 100$\%$, taxi time of 25 min and maximum flight altitude of 39,000 ft).

\begin{figure*}
	\centering
		\includegraphics[scale=.53]{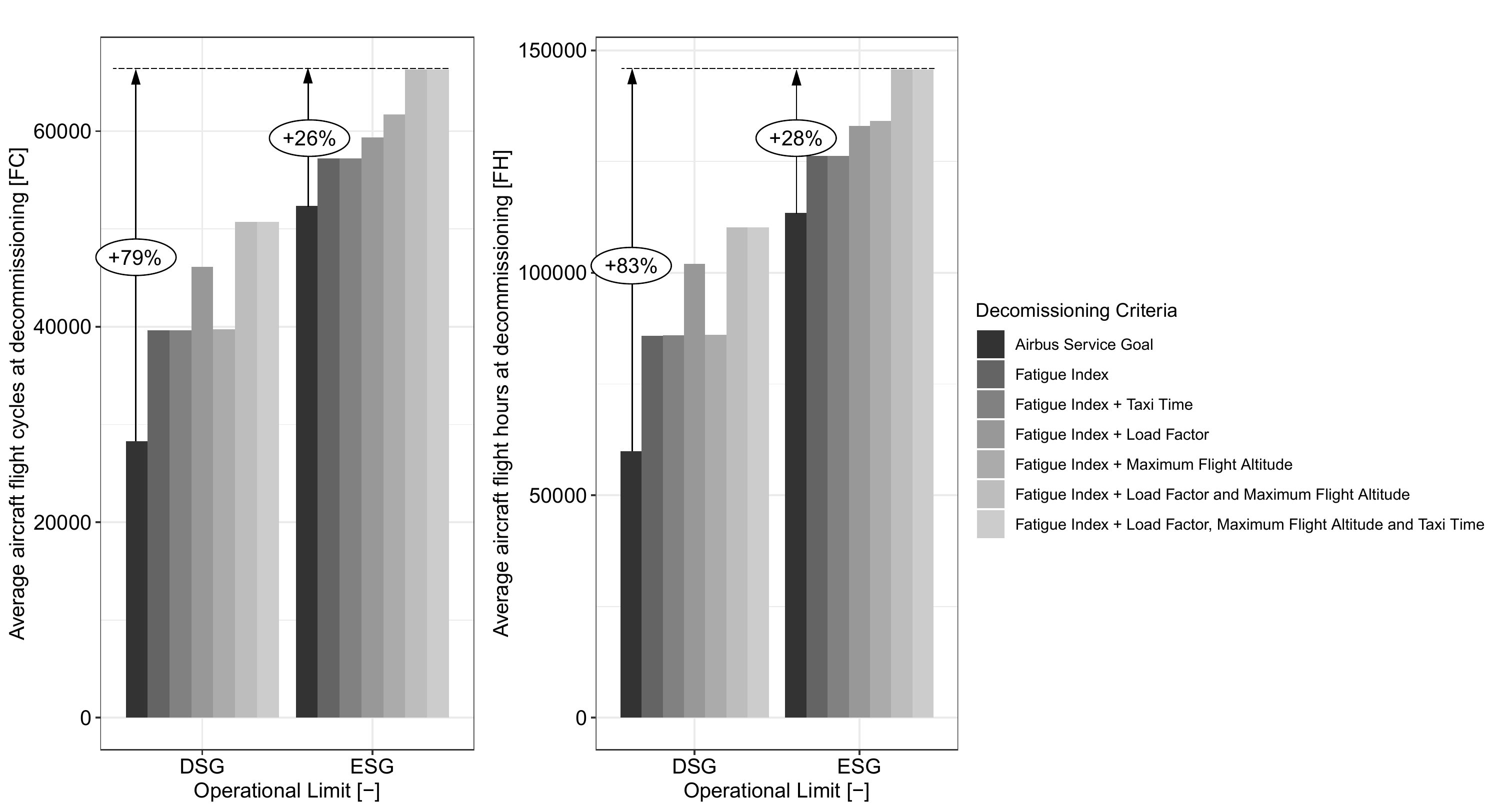}
	\caption{Computed lifetime of an average aircraft.}
	\label{fig:Lifetime}
\end{figure*}

The results shown in the following figures indicate the impact of the FDI for the Airbus A320 fleet by taking already monitored data into account. Moreover, each figure points out the extended life based on the FDI. Fig. \ref{fig:FDIs} shows the computed FDIs for different parameters. The gray and black points indicate the FDIs of an aircraft at the time of retirement due to the limit of validity given in flight cycles and flight hours for DSG and ESG, respectively. However, based on the proposed FDI, the aircraft could have been used longer as indicated by the gaps between the gray and bright red points and the black and dark red points. The bright and dark red points indicate the retirement of aircraft based on the FDIs for the DSG ($\text{FDI}_{DSG}=1$) and the ESG ($\text{FDI}_{ESG}=1$), respectively. The histograms at the edges of the diagram visualize the distributions of retirement.

Fig. \ref{fig:FDIs} shows the computed FDIs at fleet level. The scatter plots a) to f) show the FDIs of aircraft at end of life using different decommissioning criteria. All computed FDIs are based on flight cycles, flight hours, and flight distance. For the DSG, the large bright red peak in the histogram based on the wing FDI indicates that aircraft are retired almost only because of the wing. This is due to the fact that on average, aircraft fly longer than the 1.25 h per flight of the DSG. This problem was improved by introducing the ESG indicated by the dark red bars. In this case, aircraft are retired based on criticality of both indices. Comparing this to Fig.$\,$\ref{fig:FDIs}b, where the monitored taxi time is considered, no significant change can be seen. As explained in Section 3.3, this is because the taxi load's mean stress is negative and the amplitude stress is rather small. Fig. \ref{fig:FDIs}c visualizes the FDIs for considering the maximum flight altitude. This increases the effect shown in Fig. \ref{fig:FDIs}a since the considered lower flight altitude results in smaller differential pressure and therefore in a smaller fuselage FDI. Last, monitoring the seat load factor considers the known (usually lower) mean stress of the wing and therefore leads to a smaller wing FDI (see Fig. \ref{fig:FDIs}d). Aircraft are therefore retired primarily because of the fuselage, indicated by the peak of the red histograms. Fig. \ref{fig:FDIs}e and \ref{fig:FDIs}f show the fatigue indices considering the seat load factor and the flight altitude. In the latter figure, the taxi time is additionally taken into account. However, the comparison shows again that there is no significant change due to the taxi time.

Fig. \ref{fig:Lifetime} and Table \ref{Appendix_1} show the lifetime of an average aircraft in flight cycles and flight hours, where the black bars indicate decommission at the Airbus service goal. Using the FDI as the retirement criterion significantly increases the lifetime of an aircraft. This effect is demonstrated by taking the available data into account. The comparison between both Airbus service goals shows that the relative extended lifetime based on the FDI is lower in case of the ESG. However, this can be seen as a plausibility check of the applied method since this emphasizes that an aircraft structure can endure longer than the DSG initially allowed. Still, by using an FDI that takes the maximum flight altitude and seat load factor into account, aircraft to be decommissioned at ESG could be operated on average 13,941.90 flight cycles and 32,236.62 flight hours longer.

\begin{table*}[t]
	\caption{Average aircraft lifetime in flight cycles and flight hours depending on the prevailing decomissioning criterion.}
	\label{Appendix_1}
	\centering
	\begin{tabularx}{16.5 cm}{lXXXX}
		\hline
		\textbf{Decommissioning criterion} & \multicolumn{2}{l}{\thead[l]{\textbf{Average FC at} \\ \textbf{decommissioning}}} & \multicolumn{2}{l}{\thead[l]{\textbf{Average FH at} \\ \textbf{decommissioning}}} \\
		\hline
		\multicolumn{3}{l}{\textbf{Design Service Goal as Certification Limit}} \\ 
		Airbus Design Service Goal & \multicolumn{1}{r}{28,285} & $100\,\%$ & \multicolumn{1}{r}{59,949} & $100\,\%$  \\
		Fatigue Index & \multicolumn{1}{r}{39,612} & $140\,\%$ & \multicolumn{1}{r}{85,845} & $143\,\%$  \\
		Fatigue Index + Taxi Time											& \multicolumn{1}{r}{39,640} & $140\,\%$	& \multicolumn{1}{r}{85,901} & $143\,\%$  \\
		Fatigue Index + Maximum Flight Altitude								& \multicolumn{1}{r}{39,754} & $141\,\%$	& \multicolumn{1}{r}{86,035} & $144\,\%$  \\
		Fatigue Index + Load Factor & \multicolumn{1}{r}{46,119} & $163\,\%$ & \multicolumn{1}{r}{102,004} & $170\,\%$ \\
		Fatigue Index + Load Factor and Maximum Flight Altitude & \multicolumn{1}{r}{50,720} & $179\,\%$  & \multicolumn{1}{r}{110,179} & $184\,\%$ \\
		Fatigue Index + Load Factor, Maximum Flight Altitude and Taxi Time	& \multicolumn{1}{r}{50,728} & $179\,\%$ 	& \multicolumn{1}{r}{110,197} & $184\,\%$ \\ \\
		\multicolumn{3}{l}{\textbf{Extended Service Goal as Certification Limit}} \\
		Airbus Extended Service Goal													& \multicolumn{1}{r}{52,385} & $100\,\%$ 	& \multicolumn{1}{r}{113,478} & $100\,\%$ \\
		Fatigue Index & \multicolumn{1}{r}{57,202} & $109\,\%$	& \multicolumn{1}{r}{126,265} & $111\,\%$ \\
		Fatigue Index + Taxi Time											& \multicolumn{1}{r}{57,208} & $109\,\%$	& \multicolumn{1}{r}{126,283} & $111\,\%$ \\ 
		Fatigue Index + Maximum Flight Altitude								& \multicolumn{1}{r}{61,735} & $118\,\%$	& \multicolumn{1}{r}{134,116} & $118\,\%$ \\
		Fatigue Index + Load Factor											& \multicolumn{1}{r}{59,373} & $113\,\%$	& \multicolumn{1}{r}{132,971} & $117\,\%$ \\
		Fatigue Index + Load Factor and Maximum Flight Altitude				& \multicolumn{1}{r}{66,327} & $127\,\%$	& \multicolumn{1}{r}{145,715} & $128\,\%$ \\
		Fatigue Index + Load Factor, Maximum Flight Altitude and Taxi Time	& \multicolumn{1}{r}{66,329} & $127\,\%$	& \multicolumn{1}{r}{145,722} & $128\,\%$ \\
		\hline	
	\end{tabularx}
\end{table*}

\section{Discussion}
\noindent It is demonstrated that the usable lifetime of the Airbus A320 can be increased significantly without expensive fatigue tests or additional monitoring equipment. The study relies on readily available databases as well as existing approaches described in literature. Additional assumptions such as the mean stress of a wing root, the materials and their corresponding S-N curves or the sheet thickness are made. Therefore, the computed values are uncertain. Furthermore, the S-N curve describes the lifetime in a probabilistic manner. The number of flights until failure is therefore uncertain, which is not considered in this study. However, this paper makes it plausible that the proposed decommissioning criterion can extend the lifetime of aircraft.

Nevertheless, more accurate data accessible by airlines and manufacturers can improve the quality of the results. Therefore, the proposed method shall not be an exact computation of the fatigue life of aircraft rather than providing a motivation to reconsider the current definition of the limit of validity. In this line, it shall motivate improved guidelines that consider the individual usage history of aircraft in greater detail, enabling increased service goals and improved maintenance efficiency.

Moreover, all aircraft are considered to be equal with respect to variants and versions. Aircraft weight may be influenced if two or three seat classes are available, which has an impact on the weight and thus fatigue damage. However, this example emphasizes that the proposed method is easily extendable considering additional factors of influence. Also, fatigue damage computations depend on the materials used and on the design modifications, usually occurring throughout the production of an aircraft. Therefore, the computation of the FDI is specific to each version of the aircraft.

\section{Summary}
\noindent In the work at hand, a new aircraft retirement criterion based on the FDI is suggested and evaluated on fleet level. It represents a shift of the retirement criterion from flight cycles and flight hours to an Index-based criterion. The proposed index considers operational usage parameters already being monitored today, including, but not limited to, top of descent, seat load factor and taxi times, to enable increased aircraft service times. It is found that not all operational parameters affect the proposed FDI equally, with taxi times being a minor driver of fatigue. By using the FDI retirement criterion, it is possible to compare design loads with actually experienced loads of an aircraft during operation resulting in significant lifetime extensions. Given the DSG, the use of the FDI can increase the aircraft usage by an average of 79$\%$ in flight cycles and 83$\%$ in flight hours. For the ESG, average lifetime increases of 26$\%$ in flight cycles and 28$\%$ in flight hours are found. In summary, the usage of an FDI enables a more customized aircraft decommissioning decision, without changing the aircraft, impacting its required level of safety and introducing new monitoring equipment.

\section{Conclusion}
\noindent While the limit of validity of engineering data is justified and necessary to avoid the occurrence of widespread fatigue damage in airframes, it stipulates a predetermined aircraft service goal for the entire fleet. Lifetime extensions are already possible today, but they can be achieved only by additional full scale fatigue testing on the part of the manufacturer. For the Airbus A320, an extension of the DSG to the ESG, increases the number average number of flight cycles in the fleet by 85~percent and the number of flight hours by 89 percent. Defining the limit of validity as a fatigue index representing the actual state of fatigue rather than a certain amount of flight cycles and flight hours, lifetime limits can be set for individual airframes rather than entire fleets. To this end, key usage depending drivers of such a fatigue index are described and analyzed in this study. Using this approach, comparable airframe lifetime increases of 79 percent in flight cycles and 83 percent in flight hours can be achieved, without the need of additional full scale fatigue tests. Based on the results it is motivated that shifting to a more general definition of the limit of validity allows for a substantial increase in aircraft service lifetimes of individual aircraft without affecting safety and requiring additional full scale fatigue tests.

\section*{Acknowledgment}
\noindent We thank Dr.$\,$Kai-Daniel B\"uchter for reading earlier drafts and his valuable feedback. 

This research was funded by the Federal Ministry for Economic Affairs and Energy based on a decision by the German Bundestag in the national aeronautical research program LuFo V as a part of the research project Strubatex. The authors declare no conflict of interest.

\begin{figure}[h]
	\centering
	\includegraphics[scale=.12]{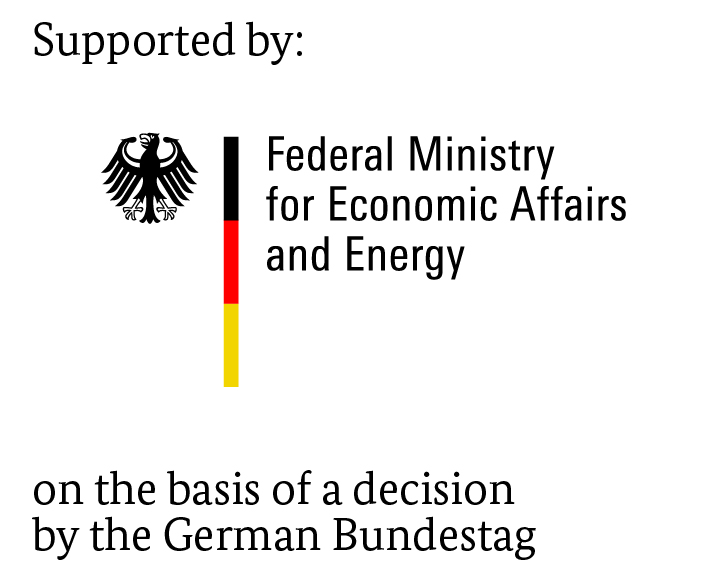}
	\label{fig:BMWi}
\end{figure}

\bibliography{bibrefNew}

\end{document}